\documentclass[12pt,preprint]{aastex}

\shorttitle{GRB Lightcurves}
\shortauthors{Fryer et al.}

\def \nuc#1#2{\relax\ifmmode{}^{#1}{\protect\text{#2}}\else${}^{#1}$#2\fi}

\begin{document}

\title{Light Curve Calculations of Supernovae from Fallback Gamma-Ray 
Bursts}

\author{Chris L. Fryer\altaffilmark{1,2}, Aimee
L. Hungerford\altaffilmark{1}, Patrick A. Young\altaffilmark{3,4}}
\altaffiltext{1}{Computational Methods, Los Alamos National Laboratories,
Los Alamos, NM 87545}
\altaffiltext{2}{Physics Dept., University of Arizona, Tucson AZ
85721}
\altaffiltext{3}{Applied Physics, Los Alamos National Laboratories,
Los Alamos, NM 87545}
\altaffiltext{4}{Steward Observatory, University of Arizona, Tucson AZ
85721}

\email{fryer@lanl.gov, aimee@lanl.gov, payoung@lanl.gov}

\begin{abstract}

The currently-favored model for long-duration gamma-ray bursts (GRBs)
invokes explosions from the collapse of a massive star down to a black
hole: either directly or through fallback.  Those GRBs forming via
fallback will produce much less radioactive nickel, and hence it has
been argued (without any real calculation) that these systems produce
dim supernovae.  These fallback black-hole GRBs have been recently
been argued as possible progenitors of a newly discovered set of GRBs
lacking any associated supernovae.  Here we present the first ever
radiation-hydrodynamics calculations of the light-curves produced in
the hypernova explosion by a delayed-fallback gamma-ray burst.  We 
find that the bolometric light-curve is dominated by shock-deposited 
energy, not the decay of radioactive elements.  As such, observations 
of such bursts actually probe the density in the progenitor wind more 
than it does the production of radioactive nickel.

\end{abstract}


\section{Introduction}

It is now generally accepted that long-duration GRBs are associated
with bright supernova-like explosions.  One of the primary differences
between these GRB-associated ``supernovae'' and normal supernovae is
that GRB-associated supernovae have larger (by more than a factor of 4
in some cases) $^{56}$Ni yields.  How is this extra $^{56}$Ni
produced.  It is known that the GRB jets themselves do not produce
large amounts of $^{56}$Ni and researchers focused on two different
paradigms for producing large amounts of $^{56}$Ni: the $^{56}$Ni is
produced in a disk around the black hole (see Surman et al. 2006 and
references therein), or the $^{56}$Ni is produced in the explosion of
the star that accompanies the GRB jet (e.g. Nagataki et al. 2003;
Mazzali et al. 2006).  Calculations assuming an accretion-disk site
for $^{56}$Ni production tend to approximate the disk using the
structure from advection dominated accretion flow solutions
(e.g. Popham et al. 1999) and use a neutrino-driven wind solution to
derive particle trajectories.  Depending upon the exact conditions in
the disk (in particular, the electron fraction), a large amount of
$^{56}$Ni can be produced.  Although Surman et al. (2006) have 
shown that the amount of $^{56}$Ni produced does depend upon the 
conditions in the disk, one can reasonably assume that this 
mechanism always produces large amounts of $^{56}$Ni.

The explosive nucleosynthesis site uses strong shocks in the stellar
explosion (beyond the narrow jet region) to heat material and cause it
to burn to $^{56}$Ni.  It depends sensitively on the structure of the
star and the velocity of the shock beyond the jet itself.  Nomoto and
collaborators \citep{Mae03,Nom04} have argued that a significant
portion of the star in a GRB is still ejected in faster-than-normal
velocities, allowing the production of enhanced amounts of $^{56}$Ni.
The problem with this site of $^{56}$Ni production is that GRBs may be
produced both as direct collapse black holes or through weak
supernovae that do not impart enough energy onto the star to prevent
the fallback of matter onto the proto-neutron star, causing it to
collapse to a black hole.  Fryer et al. (2006) found that if the delay
between this weak supernova explosion is more than a few seconds, the
amount of $^{56}$Ni produced in explosive nucleosynthesis is very
small indeed.  And with very little $^{56}$Ni produced in the
explosion, these GRBs should produce dim supernovae.

Where are these GRBs with faint associated supernovae?  All
observations suggested that bursts were associated with bright, high
$^{56}$Ni-yield, high-velocity supernovae (also known as
``hypernovae'' - Nomoto et al. 2004).  Could it be that fallback black
holes do not produce GRBs?  The same month the Fryer et al. (2006)
paper was accepted, the first of two very peculiar long-duration
bursts exploded.  These two bursts were well-localized, but their
late-time emission showed no presence of an associated supernova
(Fynbo et al. 2006, Gal-Yam et al. 2006).  On the surface, these
bursts seem to be exactly the dim-Supernova GRBs predicted by theory.
But Gal-Yam et al. (2006) argue that GRB060614 is not a typical long
burst and may require a different engine altogether.  GRB060605's
duration ($T_{90}=4 \pm 1$\,s) lies on the border between short and
long bursts and Ofek et al. (2007) argue that it is a short burst.  So
the observational evidence of low $^{56}$Ni-yield GRB-supernovae is
far from conclusive.

In addition, we must be wary of making any strong claims between
$^{56}$Ni yields and supernova light-curves.  The process by which the
energy from the decay of $^{56}$Ni and its daughter product $^{56}$Co
is converted into optical photons is not linear.  The energy from
decay is released in gamma-rays which then scatter and ultimately
absorb in the star.  The star thermalizes this energy and re-emits it
in optical photons.  But the star is also hot because it just exploded
and the shock energy has heated it.  In supernovae, the peak of
the supernova light-curve can be powered either by this shock energy
or, for the most compact stars, the decay of radioactive material.  In
this paper, we provide the first radiation-hydrodynamics explosion
calculations of these delayed-collapse hypernovae.  In \S 2, we
describe both our initial stellar models and a description of our
light-curve code.  We conclude with a discussion of our results and
comparison of these results to the current data.

\section{Initial Conditions and Code Description}

For initial conditions, we use the two 40\,M$_\odot$ fallback GRBs
from \citep{FYH06} with two different delays between the weak
supernova explosion and the GRB outburst: a 1\,s delay producing
0.33\,M$_\odot$ of $^{56}$Ni, and a 6.8\,s delay producing
0.016\,M$_\odot$.  We will focus on the longer delay explosion in our
efforts to produce a dim supernova explosion.  At early times, the
radiation is completely trapped, so we grab the models from these
calculations after the shock has moved a few times $10^9$\,cm.
\cite{FYH06} calculated the detailed yields from these models, but for
our opacities we have reduced these yields to 30 base elements from
hydrogen up to zinc.  Figure~\ref{fig:init} shows the initial velocity
structure used in our models.  Note that these explosions are very
fast with shock velocities $v_{\rm shock}$ in excess of $2 \times 10^9
{\rm \, cm s^{-1}}$, and the stars are rather compact ($1.5 \times
10^{10}$\,cm) so the shock will quickly break out of the star.

On top of this basic stellar structure, we have added a wind structure
using the mass loss rates from the last $\sim$10,000-100,000y in the
life of our Wolf-Rayet stars based on models presented by
\cite{FRY06}: $2 \times 10^{-5} {\rm M_\odot y^{-1}}$ with velocities
in the 700\,$\rm km \, s^{-1}$ range.  Note that the wind velocities
could be as high as 3,000\,$\rm km \, s^{-1}$ and it is likely that
GRBs arise from lower metallicity objects where the mass-loss rate is
lower than our chosen values.  The primary effect of both of these is
a lowering of the density in the wind.  We have included a run where
the density is scaled down by a factor of 100 to study this effect.
The initial density profile of all our models are shown in
fig.~\ref{fig:init}.

For our calculations, we use the multidimensional
radiation-hydrodynamics code RAGE (Radiation Adaptive Grid Eulerian),
which was designed to model a variety of multimaterial flows
\citep{Bal96}.  The conservative equations for mass, momentum, and
total energy are solved through a second-order, direct-Eulerian
Godunov method on a finite-volume mesh \citep{Git07}.  It includes a
flux-limited diffusion scheme to model the transport of photons using
the Levermore-Pomraning flux limiter \citep{LP91}.  RAGE has been
extensively tested on a range of verification problems
\citep{Hol99,Hue05} and applied to (and tested on) a range of
astrophysics problems~\citep{Her06,Cok06,Fry07}, including the strong
velocity gradients that exist in supernova explosions~\citep{LR06}.

The RAGE code can be used in 1,2, and 3 dimensions with spherical,
cylindrical and planar geometries in 1-dimension, cylindrical and
planar geometries in 2-dimensions, and planar geomeotries in
3-dimensions.  For this paper, we focus on 1-dimensional, spherical
calculations.  RAGE uses an adaptive mesh refinement technique,
allowing us to focus the resolution on the shock and follow the shock
as it progresses through the star.  Even so, we were forced to regrid
in our calculations to ensure that we resolved the shock at early
times but still allow us to model the shock progression out to 100\,d
(the shock moves from $10^9$\,cm out to $10^{16}$\,cm in the course of
a simulation).

For our current set of calculations, the energy released from the 
decay $^{56}$Ni and $^{56}$Co is deposited directly at the location 
of the $^{56}$Ni using the following formula:
\begin{equation}
dE/dt = E_{\rm Ni}/\tau_{\rm Ni} e^{-t/\tau_{\rm Ni}} +
E_{\rm Co}/(\tau_{\rm Co} - \tau_{\rm Ni}) 
[ e^{-t/\tau_{\rm Co}} - e^{-t/\tau_{\rm Ni}} ]
\end{equation}
where $E_{\rm Ni}=1.7$\,MeV and $E_{\rm Co}=2.9$\,MeV are the mean
energies released per atom for the decay of $^{56}$Ni and $^{56}$Co,
respectively, and $\tau_{\rm Ni}=7.6 \times 10^5$\,s, $\tau_{\rm
Co}=9.6 \times 10^6$\,s.  Especially at late times, this energy is not
deposited into the matter surrounding it, but rather escapes the star.
We have included a simulation where after the shock has reached
$5\times10^{12}$\,cm, we assume 99\% of the energy escapes, allowing a
lower bound on our models.  This provides us with 4 models in total:
1\,s delay, 6.8\,s delay, 6.8\,s delay with large escape fraction of
gamma-rays, and a 6.8\,s delay with both a large escape fraction and a
lowered wind.

For opacities, we use the SESAME opacities produced at LANL
\citep{mag95}.  These opacities have been extensively used in
astrophysics including many problems in supernovae (e.g. Fryer et
al. 2000; Deng et al. 2005; Mazzali et al. 2006).  This opacity base
is being continually updated, and we use the latest updates on these
opacities.  The assumption with these opacities is that the atoms are
in local thermodynamic equilibrium with the matter and, hence, the
opacity can be determined assuming a single temperature in each cell.
Only a few atoms deviate significantly from their local thermodynamic
solution and this affect should be minor for our calculations of
bolometric light curves.

\section{Theoretical Light Curves and the Observations}

\begin{figure}
\plotone{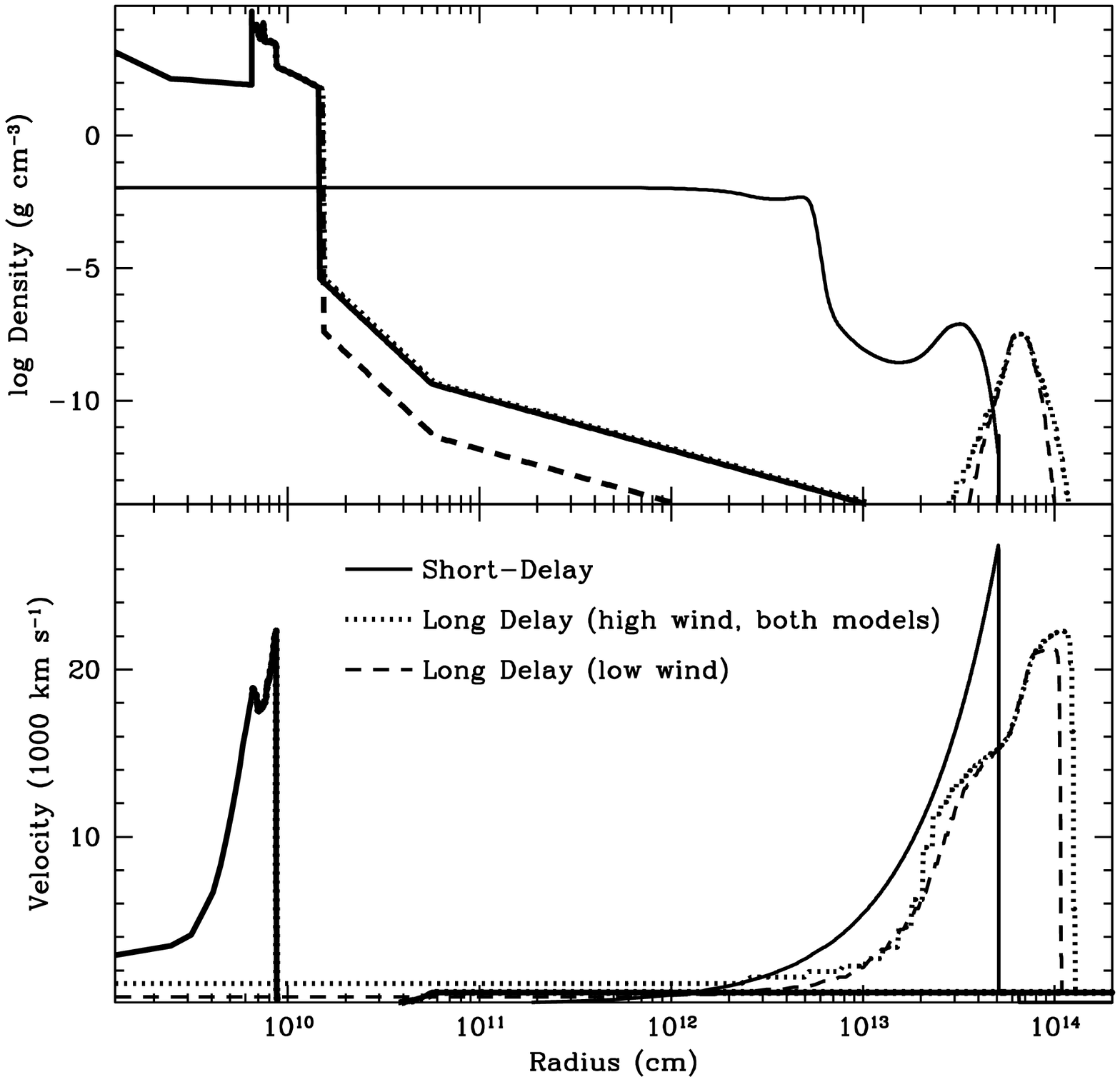}
\caption{Density versus radius (top) of our explosion models
when we map the model into RAGE and after $4 \times 10^4\,{\rm s}$ ($3
\times 10^4\,{\rm s}$ for our short delay model).  The solid line
corresponds to our short-delay model, the dotted line to our
long-delay models (the model with full energy deposition and the model
assuming 99\% of the energy escapes look very similar at this point),
and the dashed line corresponds to our long delay model with the
lowered winds.  Note that density has a peak near the shock.  This is
because most of the matter is moving at the same, very-high velocity.
The bottom panel shows the velocity versus radius for these same
models.  The short-delay model has accelerated as it expands.  At
these early times the long-delay models have not yet developed into a
homologous outflow.}
\label{fig:init}
\end{figure}

If there still is considerable internal energy in the shock when it
reaches the edge of the star, the shock can accelerate even further
when it reaches the wind.  From figure~\ref{fig:init}, we see this is
the case for our short delay model, but not for our long delay models.
At this time, roughly $4 \times 10^4\,{\rm s}$ ($3 \times 10^4\,{\rm
s}$ for our short delay model) the shock is just becoming optically
thin.  This occurs between $5-10\times 10^{13}\,{\rm cm}$ when the
shock is well beyond the stellar surface and out in the Wolf-Rayet
wind.  For our short-delay model, the shock has taken on a homologous
velocity ($v \propto r$) structure, but the long-delay models take
more than $10^5\,{\rm s}$ to develop the canonical homologous velocity
structure.  More important, note the density structure in these
models.  In the long-delay models, the weak supernova shock combined
with the accretion of fallback material produces an evacuated region
around the compact remnant.  When the strong explosions sweeps up the
weak supernova shock, it produces a shell of fast moving material.
This ring of material causes the shock to remain optically thick
longer than one would expect from more standard shock
profiles\footnote{Note that at these early times, the short-delay
models with and without the loss of $^{56}$Ni decay gamma-rays are
identical.}.  Clearly, the nature of the explosion will have a strong
affect on our light-curves.  Simple, homologous-velocity profiles,
with power-law density profiles will not be able to model the broad
range of explosion scenarios.

\begin{figure}
\plotone{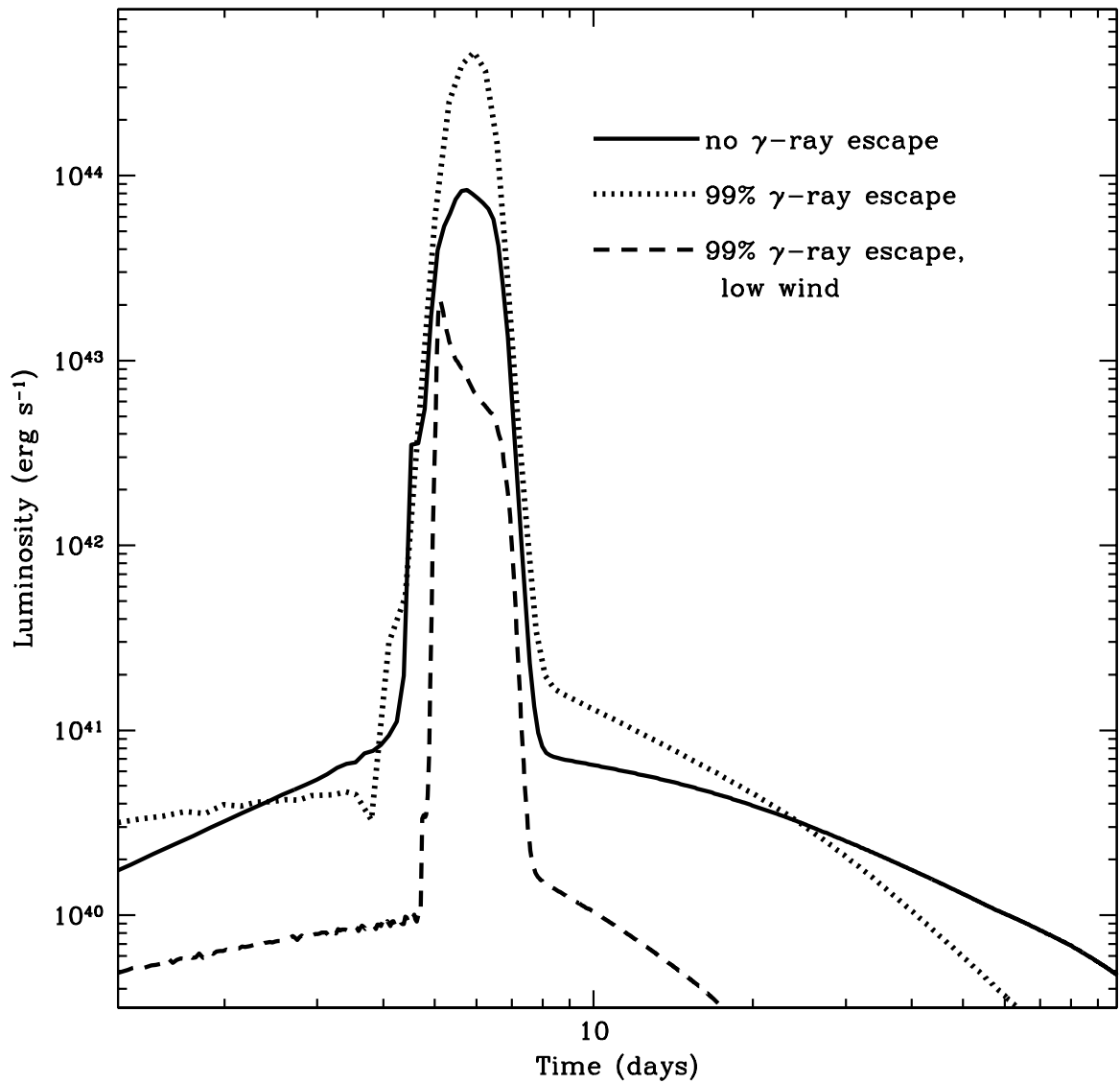}
\caption{Luminosity versus time for our 3 long-delay
simulations: full energy deposition (solid), 99\% energy loss
(dotted), and the low-wind model with 99\% energy loss (dashed).  The
peak in the light-curve is powered by shock breakout.  It happens
early (6\, days) and decays considerably by 8\, days.  By 10\,days,
the bolometric luminosity for the low wind model is below $10^40 {\rm
ergs s^{-1}}$.  The peak in the light curve is not sensitive to the
$^{56}$Ni yield, but to the energy of the explosion and the mass in
the wind.}
\label{fig:lc}
\end{figure}

The short-delay explosion experiences both a shock break-out bump at
3\,days followed by a $^{56}$Ni-powered peak at roughly 25\,days after
the launch of the explosion.  The only surprise in this model is that
shock break-out for these strong explosions propagating through dense
wind material can produce strong emission, in excess of $10^{44} {\rm
ergs \, s^{-1}}$!  This emission becomes the dominant emission in
models where the $^{56}$Ni is small.  The light-curves produced by our
long-delay ($\equiv$ low $^{56}$Ni yield) simulations are shown in
figure~\ref{fig:lc}.  These light-curves only get a large peak at
shock break-out.  This sharp peak drops within 8 days of the launch of
the explosion.  By allowing the gamma-rays to escape, we actually get
a brighter peak luminosity (the shock is slightly more compact and is
hotter), but by 20 days, the model with 100\% gamma-ray deposition is
brighter and by 100\,d it has a luminosity that is an order of
magnitude higher than the model with only 1\% gamma-ray deposition.

These features can be explained if the bulk of the light-curve is
powered by shock energy, not gamma-ray deposition.  The peak in the
light curve is powered by shock break-out.  Indeed, it appears that
shock energy dominates out beyond 20-30\,days.  The fact that the
supernova explosion is strong ($\gtrsim 10^{52}$\,erg) and the
$^{56}$Ni production is low ($<0.02\,M_\odot$), this is not 
unexpected.  In such a system, one would expect that the amount 
of mass in the wind is more important.  Our simulation with a 
low wind is an order of magnitude dimmer than our strong wind 
candidates.

The strong shock drives the matter temperature (and the emitted
Planck-averaged radiation temperature) well beyond the optical bands.
But as the shock moves out of the envelope, the averaged radiation
energy in the shock decreases (Fig. 3).  It is the temperature in 
the shocks that drive most of the emission and this temperature 
is characteristic of the radiation temperature.  6 days into the explosion,
when the luminosity peaks, the Planck-averaged radiation temperature
lies between 1.5-4eV for all models, with most of the photons 
emerging in the ultraviolet.  

Let's compare these models with the observational limits placed by
Fynbo et al.(2006) and Gal-Yam et al. (2006).  At 10 days for a GRB at
388\,Mpc (roughly equivalent to a redshift of 0.089 corresponding to
GRB060505), the bolometric luminosities for these three models are
23.5, 24.7, and 26.8 magnitudes for the 1\% gamma-ray deposition,
100\% gamma-ray deposition, and low wind models respectively.  For the
more distant GRB060614, the expected late-time flux is even lower.  It
is likely that these models would all be below the observed limits at
this time.

However, at shock break-out, these spherically-symmetric models
predict a very bright luminosity (above 20th magnitude at 6\,d).  Why
don't we see the burst at this time?  Although the afterglow is still
strong at this time, our predicted luminosities could well dominate.
But remember that the peak flux is in the ultraviolet, not the
optical.  The fact that nothing is observed could place constraints on
the model.  Either there is considerable dust extinction (a few
magnitudes in the UV), the explosion is weaker than our spherically
symmetric $10^{52}$\,erg explosion has assumed, or this GRB truly is
different than our standard long-duration burst classification as
Gehrels et al. (2006) and Gal-Yam et al. (2006) argue.

We have learned a number of features of delayed black-hole forming
GRBs.  First, the emission of these supernovae are likely to peak at
shock break-out.  The luminosity is dominated by the shock energy, and
not the decay of radioactive elements.  As such, any observation
places more constraints on the surrounding wind than it does on the
$^{56}$Ni yield.  Also, the structure of the shock may well be very
different than the canonical density and velocity structures seen in
more normal supernovae and we must pay particular attention to the
explosions in calculating these light-curves.  The models in these
papers are examples of a larger set of possibilities.  Finally, bear
in mind that we are assuming spherically symmetric models run in the
gray approximation and asymmetric, multi-group models may have very
different light-curves.

\acknowledgements This work was carried out in part under the auspices
of the National Nuclear Security Administration of the U.S. Department
of Energy at Los Alamos National Laboratory and supported by Contract
No. DE-AC52-06NA25396.

\clearpage

\clearpage

\clearpage

\begin{figure}
\plotone{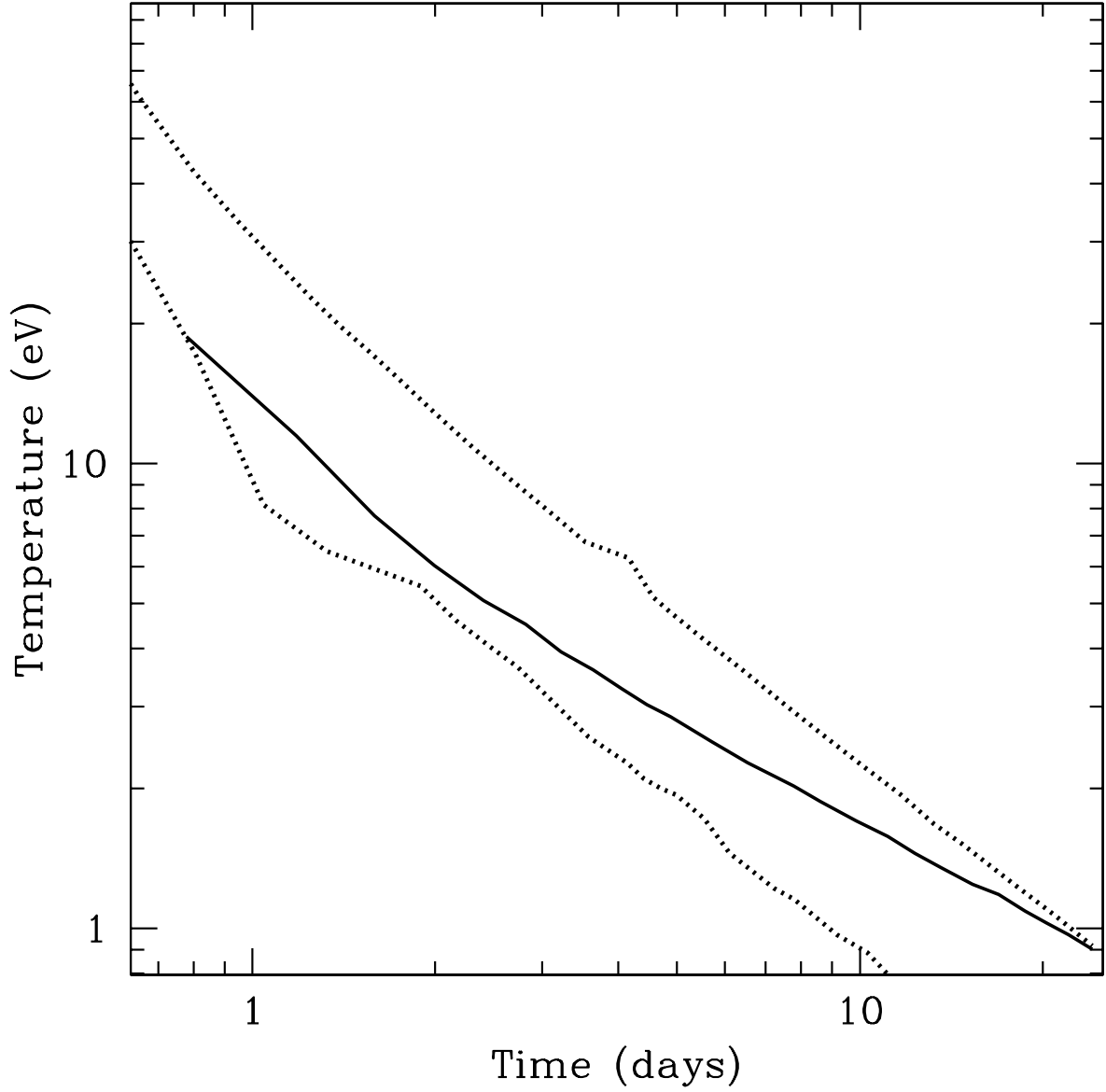}
\caption{Planck-averaged temperature of the radiation in the shock as
a function of time.  Recall that the peak in the emission occurs at a
wavelength at $\lambda_{\rm peak} \approx 250/T_{\rm eV}$nm where
$T_{\rm eV}$ is the temperature in eV.  The peak in the luminosity
occurs at roughly 6 days for all of our models, and the bulk of the 
emission is radiated in the ultraviolet.}
\label{fig:temp}
\end{figure}
\clearpage

\end{document}